\documentclass[a4paper,11pt]{article}
\usepackage{jcappub} 
\usepackage{lineno}

\usepackage{graphicx}
\usepackage{dcolumn}
\usepackage{bm}
\usepackage{amssymb}
\usepackage{latexsym}
\usepackage{booktabs}
\usepackage{amsmath}
\usepackage{multirow}
\usepackage{url}
\usepackage{hyperref}
\usepackage{footnote}
\usepackage{float}
\usepackage{threeparttable}
\usepackage{color}
\usepackage{array}
\usepackage{enumerate}
\usepackage{adjustbox}

\usepackage{makecell}
\usepackage{diagbox}
\usepackage{epstopdf}
\usepackage{epsfig}
\usepackage{longtable}
\usepackage{supertabular}
\usepackage{algorithm}
\usepackage{pifont}
\usepackage{algorithmic}
\usepackage{changepage}
\usepackage{setspace}

\pdfoutput=1

\begin{document}

\title{Joint constraints on cosmic birefringence and early dark energy from ACT, Planck, DESI, and PantheonPlus}

\author[a,\ast]{Lu Yin,}
\author[b]{Guo-Hong Du,}
\author[b]{Tian-Nuo Li}
\author[b,c,d,\ast]{and Xin Zhang\note[$\ast$]{Corresponding author.}}

\affiliation[a]{Department of Physics, Shanghai University, Shanghai 200444,  China.}
\affiliation[b]{Liaoning Key Laboratory of Cosmology and Astrophysics, College of Sciences, Northeastern University, Shenyang 110819, China.}
\affiliation[c]{MOE Key Laboratory of Data Analytics and Optimization for Smart Industry, Northeastern University, Shenyang 110819, China.}
\affiliation[d]{National Frontiers Science Center for Industrial Intelligence and Systems Optimization, Northeastern University, Shenyang 110819, China.}

\emailAdd{yinlu@shu.edu.cn, duguohong@stumail.neu.edu.cn, litiannuo@stumail.neu.edu.cn, zhangxin@neu.edu.cn}
\date{\today}

\abstract{
With the increasing number of high-precision astronomical observations, physical quantities that were previously inaccessible to accurate calculations, such as cosmic birefringence, have once again become a focal point of interest. Such phenomena induce a nonvanishing cross-correlation between the $E$- and $B$-mode polarizations of the cosmic microwave background (CMB), thereby providing a direct observational signature of parity violation. The Chern--Simons coupling between the scalar field in early dark energy (EDE) models and CMB photons is regarded as a plausible mechanism for generating cosmic birefringence. 
Recent data from the Atacama Cosmology Telescope (ACT) deliver $EB$ measurements at higher multipole moments than those previously achieved by {Planck}, while DESI and PantheonPlus datasets provide new and stringent constraints on the late-time expansion history. Using a joint analysis of {Planck}, DESI DR2, PantheonPlus, and ACT data, we perform a full-parameter constraint on the cosmic birefringence effects induced by the EDE-CMB photon coupling. Our results favor a higher Hubble constant, $H_0 = 76.9^{+2.9}_{-2.5}\,\rm km\,s^{-1}\,Mpc^{-1}$,
and a relatively large EDE fraction, $f_{\mathrm{EDE}} = 0.232^{+0.074}_{-0.047}$.
By comparing the cosmological evolution of this model across different data combinations, we find that the ACT-$EB$ data combined with {Planck} + DESI + PantheonPlus provide consistent constraints.
}

\maketitle

\section{Introduction}

Cosmic birefringence describes a phenomenon in which the plane of linear polarization of cosmic microwave background (CMB) photons undergoes a rotation as the photons propagate over cosmological distances \cite{Carroll:1989vb, Carroll:1991zs, Harari:1992ea}. Analyses based on Planck observations have reported evidence for a non-vanishing rotation angle $\beta$ at approximately the $3.6\sigma$ level, with a best-fit value of $\beta = {0.342^{\circ}}^{+0.094^\circ}_{-0.091^\circ}$ at $68\%$ confidence level~\cite{Minami:2020odp, Eskilt:2022cff}. Observationally, this effect manifests itself through non-trivial correlations between the $E$- and $B$-mode polarization components \cite{Lue:1998mq}.

Within the framework of parity-invariant physics, the harmonic coefficients $E_{\ell m}$ and $B_{\ell m}$ transform with definite parity under spatial inversion. As a consequence, their auto-power spectra remain unchanged, while the cross-spectrum $C_\ell^{EB}$ is required to vanish. The detection of a non-zero $EB$ correlation therefore provides a direct signature of parity violation \cite{Naokawa:2023upt, Namikawa:2023zux, Ferreira:2023jbu, Yin:2023srb, Greco:2024oie, Namikawa:2024dgj, Jamieson:2024mau, Sullivan:2025btc, LiteBIRD:2025yfb, Lonappan:2025hwz, Namikawa:2025sft, Ballardini:2025apf, Namikawa:2025doa, Diego-Palazuelos:2025dmh}, and may also arise from CPT-violating mechanisms \cite{Feng:2006dp, Li:2006ss, Li:2008tma, Xia:2008si, Xia:2007qs, Xia:2009ah, Li:2014oia} or anisotropic effects \cite{Li:2013vga, Luongo:2021nqh, Krishnan:2021dyb, LiteBIRD:2024dbi, Namikawa:2024sax}.

A well-motivated theoretical origin of cosmic birefringence involves axion-like scalar fields coupled to electromagnetism through a Chern-Simons interaction of the form $g\phi F_{\mu\nu}\tilde{F}^{\mu\nu}$ \cite{Finelli:2008jv, Choi:2021aze, Nakatsuka:2022epj, Murai:2022zur, Galaverni:2023zhv}. Such a coupling induces a rotation of the polarization angle, generates non-zero $EB$ correlations, and transfers power from $E$ modes into $B$ modes \cite{Monelli:2023wmv, ACT:2025zrv}
. At present, multiple CMB experiments have already reported measurements of the $EB$ spectrum, including POLARBEAR \cite{POLARBEAR:2019snn}, ACT \cite{ACT:2025fju, ACT:2025qjh}, SPT \cite{SPT-3G:2021eoc}, and SPIDER \cite{SPIDER:2021ncy}. Looking forward, forthcoming polarization missions such as the Simons Observatory \cite{SimonsObservatory:2018koc}, AliCPT \cite{Gao:2017cra, Li:2017drr}, and LiteBIRD \cite{LiteBIRD:2022cnt, LiteBIRD:2023aov, LiteBIRD:2023zmo, LiteBIRD:2024dbi, LiteBIRD:2024twk, LiteBIRD:2024eau, LiteBIRD:2025tnn, LiteBIRD:2025trg} are expected to dramatically improve sensitivity to birefringence effects. In particular, LiteBIRD is anticipated to be capable of detecting not only a primary $EB$ signal but also the induced secondary $BB$ contribution generated by polarization rotation.

At the same time, modern cosmology faces another major challenge in the form of the Hubble tension, which refers to the persistent discrepancy between the value of the Hubble constant ($H_0$) inferred from early-Universe probes, such as the CMB and baryon acoustic oscillation (BAO), and that obtained from local distance-ladder measurements \cite{Bernal:2016gxb, Riess:2021jrx}. This disagreement has now reached the $4$--$5\sigma$ level and has become increasingly difficult to dismiss \cite{Guo:2018ans,DiValentino:2021izs,DiValentino:2022fjm,Giare:2023xoc,Vagnozzi:2019ezj,Vagnozzi:2023nrq,Escamilla:2024ahl,Du:2024pai,Li:2024qso,Cai:2025mas,Li:2025nnk,Qiu:2024sdd, Feng:2025mlo,Smith:2025icl,Lee:2025yvn,Piras:2025eip,Li:2025owk, CosmoVerseNetwork:2025alb}. A wide range of theoretical efforts have focused on modifying the expansion history of the Universe in order to reconcile these measurements.

Among the proposed solutions, early dark energy (EDE) scenarios stand out as one of the few frameworks capable of easing the tension without introducing severe conflicts with other cosmological observations. In these models, a transient dark energy component becomes dynamically relevant prior to recombination, modifying the sound horizon at last scattering and consequently shifting the inferred value of $H_0$. EDE models have been extensively confronted with observational data \cite{Yin:2020dwl,Colgain:2021pmf,Krishnan:2021dyb,Krishnan:2021jmh,Akarsu:2022lhx,Murai:2022zur,Poulin:2018cxd,Herold:2023vzx,Efstathiou:2023fbn,Simon:2024jmu,Lin:2025gne,Giare:2024akf,Forconi:2023hsj}, yielding mixed conclusions. While some studies suggest tension between EDE and cosmic birefringence measurements \cite{Eskilt:2023nxm, Piras:2025eip}, others find that EDE remains fully compatible with both birefringence constraints and the $H_0$ determination from SH0ES \cite{Kochappan:2024jyf}. Meanwhile, recent observational advances from Dark Energy Spectroscopic Instrument (DESI)~\cite{DESI:2025zgx} and PantheonPlus~\cite{Brout:2022vxf} have the potential to significantly refine our understanding of the evolution of cosmological parameters. By providing high-precision measurements of large-scale structure and an expanded, homogeneous type Ia Supernova sample, these datasets probe both the late-time expansion history and its connection to early-Universe physics with unprecedented accuracy. The emerging hints of deviations from a strictly constant dark energy equation of state, together with tighter constraints on distance-redshift relations, suggest that parameters traditionally treated as fixed may exhibit subtle time dependence. As a result, the inclusion of DESI and PantheonPlus data opens new avenues for reassessing the consistency of the standard cosmological model and for exploring extensions in which key parameters evolve across cosmic epochs~\cite{Giare:2024oil,Giare:2024smz,Giare:2024gpk,Wang:2026kbg,Li:2026xaz,Li:2025vuh,Jiang:2024viw,Liu:2025mub,Li:2024bwr,Li:2025cxn,Li:2025muv,Du:2025xes,Du:2024pai,Wolf:2025jed,Pang:2025lvh,Paliathanasis:2026ymi,Pan:2025qwy,Yang:2025uyv,Li:2024qus,Zhang:2025dwu,Li:2026asg}.

This work investigates cosmic birefringence induced by a scalar-photon coupling embedded within an EDE framework, and places constraints on a broad set of cosmological parameters using the latest observational data from ACT, DESI,  PantheonPlus, and Planck. Such a scenario naturally links the physical origin of the Hubble tension with the mechanism responsible for polarization rotation in the CMB \cite{Lee:2002nv, Das:2013sca, Das:2023hbw}.
Our primary objective is to examine how different combinations of current observational datasets-including ACT, Planck, DESI, PantheonPlus, BOSS, and SH0ES-affect the inferred constraints on the cosmic birefringence coupling constant $g M_{\rm Pl}$, the Hubble parameter, and other cosmological parameters, and to identify the corresponding trends and dataset-dependent preferences.

This paper is organized as follows. In Section~\ref{sec:2}, we review the EDE model under consideration, present the relevant Boltzmann equations, and describe their modification in the presence of scalar-photon couplings. Section~\ref{sec:3} contains our main results, including constraints on $g M_{\rm Pl}$ and the Hubble parameter obtained from different datasets. We summarize our conclusions in Section~\ref{sec:4}.


\section{Cosmic birefringence from the EDE model }
\label{sec:2}

By augmenting the theory with a Chern--Simons interaction, the dynamics of a pseudoscalar field coupled to electromagnetism can be described by the Lagrangian~\cite{Murai:2022zur}
\begin{equation}
\mathcal{L}
= -\frac{1}{2}(\partial_\mu \phi)(\partial^\mu \phi)
- V(\phi)
- \frac{1}{4} F_{\mu\nu}F^{\mu\nu}
- \frac{1}{4} g\,\phi\,F_{\mu\nu}\tilde{F}^{\mu\nu} ,
\end{equation}
where $\phi$ denotes a pseudoscalar field endowed with a canonical kinetic term and a self-interaction potential $V(\phi)$. The parameter $g$, which carries mass dimension $-1$, quantifies the strength of the Chern--Simons coupling. Here $F_{\mu\nu}$ is the electromagnetic field strength tensor, and $\tilde{F}_{\mu\nu}$ is its dual.

EDE refers to a cosmological component that contributes a non-negligible fraction of the total energy density around the epoch of matter--radiation equality. Such a component has been proposed as a possible resolution to the Hubble tension by modifying the expansion history prior to recombination
\cite{Caldwell:2003vp, Smith:2019ihp, Berghaus:2019cls, Alexander:2019rsc, Chudaykin:2020acu, Agrawal:2019lmo, Niedermann:2019olb, Freese:2004vs, Ye:2020btb, Akarsu:2019hmw, Lin:2019qug, Yin:2020dwl, Braglia:2020bym}.
A commonly adopted form of the EDE potential is
\begin{equation}
V(\phi)=\Lambda^4\bigl[1-\cos(\phi/f)\bigr]^n ,
\label{eq:pot}
\end{equation}
where $n$ is a phenomenological parameter; in this work, we restrict our analysis to the case $n=3$ { as its constraints are in good agreement with other observations}.
{ The choice of $n=3$ is motivated by previous analyses showing that it provides a good fit to the Planck $EB$ data while remaining consistent with other CMB constraints \cite{Murai:2022zur}. In addition, recent studies comparing different potential indices within the EDE+birefringence framework indicate that $ n=3$ yields the most favorable constraints \cite{Zhang:2026fzj}.}
Within this framework, the EDE energy density becomes dynamically relevant before photon decoupling, leading to a reduction of the comoving sound horizon. The pseudoscalar nature of the potential in Eq.~(\ref{eq:pot}) allows the EDE field to simultaneously act as a source of cosmic birefringence.
{  We define the dimensionless field variable $\theta \equiv \phi/f$. This model assumes a cosmological-constant late-time dark energy component; the EDE field modifies only the pre-recombination expansion history and is negligible at late times.}

The inclusion of the Chern--Simons term modifies the propagation of electromagnetic waves, resulting in a helicity-dependent dispersion relation
\cite{Carroll:1989vb, Carroll:1991zs, Harari:1992ea}
\begin{equation}
\omega_{\pm} \simeq k \mp \frac{g}{2}
\left( \frac{\partial \phi}{\partial t}
+ \frac{\mathbf{k}}{k}\cdot\nabla\phi \right)
= k \mp \frac{g}{2}\frac{\mathrm{d}\phi}{\mathrm{d}t} ,
\end{equation}
where $\omega_{\pm}$ correspond to the angular frequencies of right- and left-handed circularly polarized photon modes, respectively. We adopt a right-handed coordinate system in which the $z$-axis is aligned with the photon propagation direction.

The helicity-dependent dispersion induces a rotation of the plane of linear polarization. In the limit where the photon frequency is much larger than the time variation rate of $\phi$, the WKB approximation applies. The accumulated rotation angle between an initial time $t$ and the present time $t_0$ is then given by
\begin{equation}
\beta(t)
= -\frac{1}{2}\int_t^{t_0} \mathrm{d}\tilde{t}\,(\omega_+-\omega_-)
= \frac{g}{2}\bigl[\phi(t_0)-\phi(t)\bigr] .
\label{beta}
\end{equation}

With our sign conventions, a positive value of $\beta$ corresponds to a clockwise rotation of the polarization direction on the sky, with the $z$-axis taken along the observer’s line of sight. The polarization angle convention follows that adopted in Ref.~\cite{Komatsu:2022nvu}.

The evolution of CMB polarization in the presence of cosmic birefringence is governed by a modified Boltzmann equation,
\begin{equation}
{}_{\pm 2}\Delta_P'
+ i q \mu\,{}_{\pm 2}\Delta_P
= \tau'\!\left[
-{}_{\pm 2}\Delta_P
+ \sqrt{\frac{6\pi}{5}}\,{}_{\pm 2}Y_2^0(\mu)\,\Pi(\eta,q)
\right]
\pm 2 i \beta'\,{}_{\pm 2}\Delta_P ,
\end{equation}
where $\eta$ is the conformal time, $q$ denotes the Fourier wave number, and $\mu$ is the cosine of the angle between the photon direction and the wave vector. The quantities ${}_{\pm 2}Y_\ell^m$ are spin-2 spherical harmonics, $\Pi(\eta,q)$ is the polarization source function, and ${}_{\pm 2}\Delta_P$ represents the Fourier transform of $Q\pm iU$, with $Q$ and $U$ being the Stokes parameters.

Expanding ${}_{\pm 2}\Delta_P$ in spin-weighted harmonics yields
\begin{equation}
{}_{\pm 2}\Delta_P(\eta,q,\mu)
= \sum_\ell i^{-\ell}\sqrt{4\pi(2\ell+1)}
\,{}_{\pm 2}\Delta_{P,\ell}(\eta,q)\,
{}_{\pm 2}Y_\ell^0(\mu) .
\end{equation}

When the rotation angle varies with conformal time, the solution for the polarization multipoles at the present epoch takes the form
\begin{equation}
{}_{\pm 2}\Delta_{P,\ell}(\eta_0,q)
= -\frac{3}{4}\sqrt{\frac{(\ell+2)!}{(\ell-2)!}}
\int_0^{\eta_0} \mathrm{d}\eta\,
\tau' e^{-\tau(\eta)}\,
\Pi(\eta,q)\,
\frac{j_\ell(x)}{x^2}
\,e^{\pm 2 i \beta(\eta)} ,
\label{Boltzmann}
\end{equation}
where $x=q(\eta_0-\eta)$ and $j_\ell(x)$ is the spherical Bessel function.

The angular power spectra of CMB polarization are given by
\begin{equation}
C_\ell^{XY}
= 4\pi\int \mathrm{d}(\ln q)\,
\mathcal{P}_s(q)\,
\Delta_{X,\ell}(q)\,
\Delta_{Y,\ell}(q) ,
\end{equation}
with $\mathcal{P}_s(q)$ the primordial scalar power spectrum and $X,Y\in\{E,B\}$ \cite{Liu:2006uh}. The $E$- and $B$-mode multipoles are obtained through
\begin{equation}
\Delta_{E,\ell}(q)\pm i\Delta_{B,\ell}(q)
\equiv -\,{}_{\pm 2}\Delta_{P,\ell}(\eta_0,q) .
\end{equation}

In the special case of a constant rotation angle, the observed polarization modes satisfy
\begin{equation}
\Delta_{E,\ell}\pm i\Delta_{B,\ell}
= e^{\pm 2 i\beta}
\left(\tilde{\Delta}_{E,\ell}\pm i\tilde{\Delta}_{B,\ell}\right) ,
\end{equation}
where tilded quantities denote the spectra in the absence of birefringence. The corresponding power spectra are
\begin{equation}
C_\ell^{EB}
= \frac{1}{2}\sin(4\beta)
\left(\tilde{C}_\ell^{EE}-\tilde{C}_\ell^{BB}\right) ,
\label{eq:cleb}
\end{equation}
When $\beta=0$, the standard $TE$, $EE$, and $BB$ spectra are recovered and the $EB$ correlations vanish. In the limit $\tilde{C}_\ell^{BB}\ll \tilde{C}_\ell^{EE}$, one finds the approximate relation
\begin{equation}
C_\ell^{EB}\simeq \tan(2\beta)\,C_\ell^{EE} .
\label{eq:cleb2}
\end{equation}

In the general case where the scalar field evolves dynamically, the rotation angle $\beta$ must be computed numerically together with the Boltzmann equations. We perform this calculation using the public \texttt{CLASS\_EDE} extension \cite{Hill:2020osr} of the \texttt{CLASS} code \cite{Lesgourgues:2011re, Blas:2011rf}.

\section{Fitting new form of cosmic birefringence with Planck and ACT observations}
\label{sec:3}

Our analysis primarily leverages data from Planck \cite{Planck:2019nip, Planck:2018lbu}, DESI~\cite{DESI:2025zgx}, and PantheonPlus~\cite{Brout:2022vxf}. The Planck satellite mission provides one of the most definitive measurements of the CMB, offering high-precision observations of temperature anisotropies and polarization power spectra ($TT$, $TE$, and $EE$). In addition, we include observations from the DESI, which is constructing an unprecedented three-dimensional map of the Universe by accurately measuring distances to tens of millions of galaxies and quasars. Notably, its data release (DR2) suggests a potential time-evolution of dark energy-a finding that, if confirmed, would significantly challenge the standard $\Lambda$CDM cosmological framework. We also employ the PantheonPlus compilation, which integrates measurements from 1,550 type Ia supernovae. \footnote{{ Although DESY5's~\cite{DES:2024jxu} supernova result provides good controlled calibrations, Refs. \cite{Efstathiou:2024xcq, DES:2025tir, Huang:2025som, DES:2025sig, DESI:2025zgx} observed a discrepancy in the standardized luminosities of low- and high-z supernovae common to the DESY5 and PantheonPlus samples. We only select PantheonPlus in this analysis.}} These results provide stringent tests of the standard model by precisely constraining the matter-energy content and the epoch of cosmic acceleration, while yielding updated determinations of the Hubble constant.

Building upon the above baseline datasets, we include Planck-$EB$~\cite{Planck:2020olo, Eskilt:2023nxm} and ACT-$EB$~\cite{ACT:2025fju} measurements to investigate the influence of $EB$ polarization observations at low and high multipoles, respectively, on the inferred cosmological parameters. For the Planck-$EB$ component, we used the spectrum analyzed by Eskilt et al. \cite{Eskilt:2022cff}, comprising 72 data points across $50 \leq \ell \leq 1500$. The ACT is a new-generation ground-based experiment, which is utilized for its exceptional sensitivity at high multipoles. We use the most recent $EB$ data release \cite{ACT:2025fju}, which includes 38 data points spanning $600 \leq \ell \leq 3400$. These high-$\ell$ $EB$ measurements are expected to provide constraints on model parameters, especially the Chern-Simons coupling constant $g M_{\mathrm{Pl}}$.

\begin{table}[htbp]
\renewcommand\arraystretch{1.5}
\centering
\caption{Cosmological parameters from Planck-$EB$ (with Planck+DESI +PantheonPlus) dataset and ACT-$EB$ (with Planck+DESI+PantheonPlus) dataset analyses.}
\resizebox{0.8\textwidth}{!}{%
\setlength{\tabcolsep}{10pt}
\begin{tabular}{|l|p{4cm}|p{4cm}|}
\hline
Parameter & Planck-$EB$ (with Planck +DESI+PantheonPlus) & ACT-$EB$ (with Planck +DESI+PantheonPlus) \\
\hline
$\ln(10^{10} A_\mathrm{s})$ & $3.288 \pm 0.019$ & $3.290^{+0.020}_{-0.017}$ \\
$n_\mathrm{s}$ & $0.979 \pm 0.013$ & $0.983^{+0.014}_{-0.013}$ \\
$H_0\,[\rm km\,s^{-1}\,Mpc^{-1}]$ & $76.2^{+3.1}_{-2.5}$ & $76.9^{+2.9}_{-2.5}$ \\
$\Omega_\mathrm{b} h^2$ & $0.02421^{+0.00054}_{-0.00060}$ & $0.02439 \pm 0.00060$ \\
$\Omega_\mathrm{c} h^2$ & $0.150^{+0.012}_{-0.011}$ & $0.153 \pm 0.011$ \\
$\tau_\mathrm{reio}$ & $0.136^{+0.012}_{-0.0080}$ & $0.135^{+0.013}_{-0.0085}$ \\
$f_{\rm EDE}$ & $0.215^{+0.083}_{-0.049}$ & $0.232^{+0.074}_{-0.047}$ \\
$gM_\mathrm{Pl}$ & $0.164^{+0.038}_{-0.079}$ & $0.129^{+0.031}_{-0.075}$ \\
$\log_{10}z_c$ & $3.554^{+0.030}_{-0.006}$ & $3.576^{+0.029}_{-0.007}$ \\
$S_8$ & $0.964^{+0.024}_{-0.021}$ & $0.967^{+0.022}_{-0.020}$ \\
\hline
$\chi^2_\mathrm{min}$ & $3195.31$ & $2982.60$ \\
\hline
\end{tabular}
}
\label{tab:cosmo_params}
\end{table}

\begin{figure*}[htbp]
\includegraphics[scale=0.8]{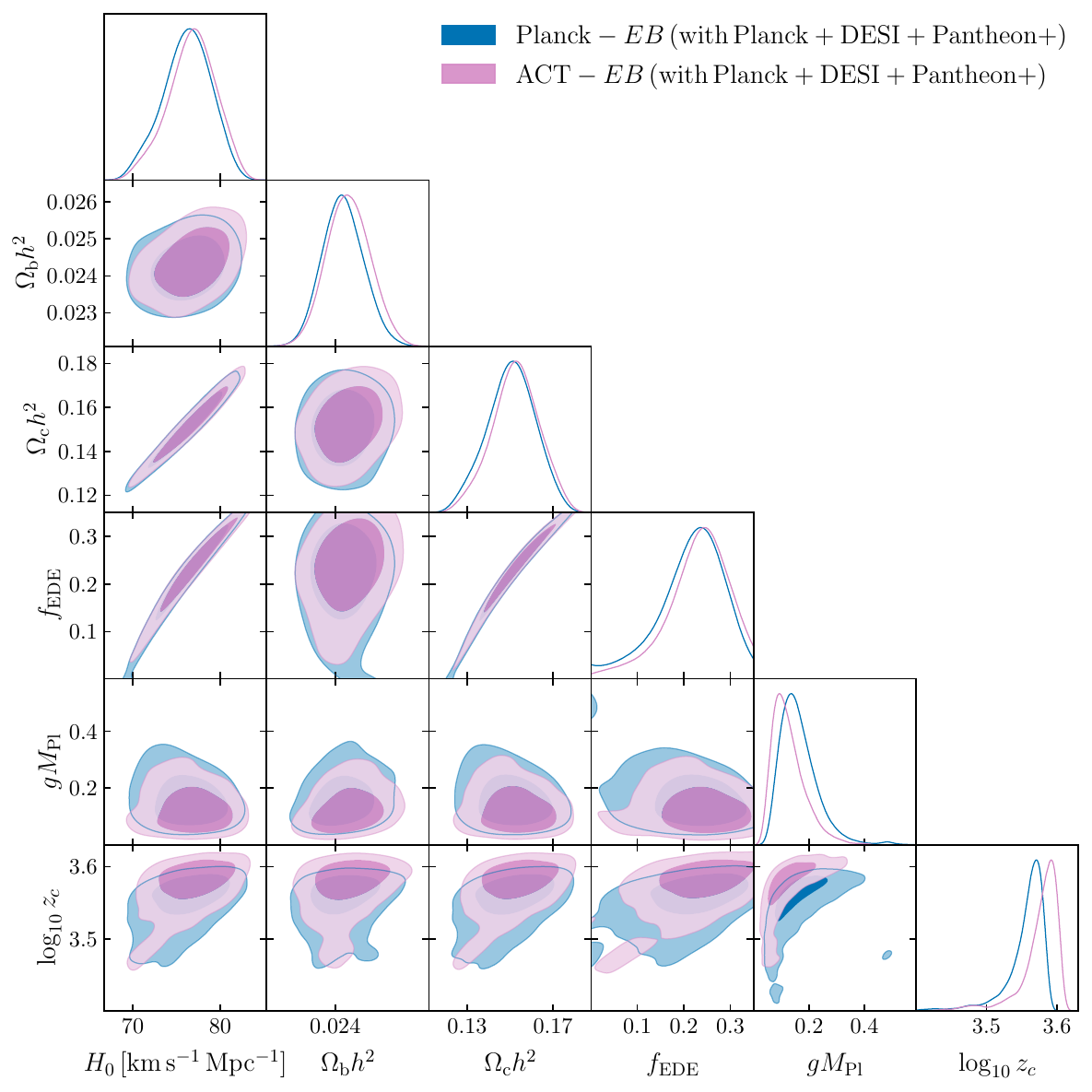}
\centering
\caption{\label{fig1} One and two-dimensional distributions of $H_0$, $\Omega_\mathrm{b} h^2$, $\Omega_\mathrm{c} h^2$, $f_\mathrm{EDE}$, $gM_\mathrm{Pl}$, and $\log_{10}z_c$, where the contour lines represent 68$\%$ and 95$\%$ C.L., respectively. The blue color contour corresponds to the fitting results from the Planck-EB with Planck+DESI+PantheonPlus dataset. The purple color corresponds to the fitting results from the ACT-$EB$ with Planck+DESI + PantheonPlus dataset. }
\end{figure*}

We define two primary data combinations: A joint analysis of the Planck-$EB$ polarization spectrum with the Planck $TT, TE, EE$ spectra, supplemented by the DESI DR2 and PantheonPlus datasets, we write the joint as Planck-$EB$ (with Planck + DESI + PantheonPlus). The second data group is a combination of the ACT-$EB$ measurements with the aforementioned Planck, DESI, and PantheonPlus; we refer to it as ACT-$EB$ (with Planck+DESI+PantheonPlus). Based on these combinations, we perform parameter inference using Markov Chain Monte Carlo (MCMC) techniques in the Cobaya program \cite{Torrado:2020dgo}. 
In our analysis, the parameter $\theta_i$ is fixed to 2.7,  { which is consistent with the best-fit $\theta_i$ found in previous EDE analyses with the $n=3$ potential \cite{Murai:2022zur, Eskilt:2023nxm}, and is adopted here as a fiducial choice to limit the dimensionality of the parameter space.} { We fix the sum of neutrino masses to $\sum m_\nu = 0.06\ \mathrm{eV}$, consistent with the Planck baseline assumption.} The best-fit values for nine free parameters are derived: the six for standard $\Lambda$CDM parameters, two for the EDE model ($f_{\rm EDE}$ and $\log_{10} z_c$), and the Chern-Simons coupling constant $g M_{\rm Pl}$ associated with cosmic birefringence. The resulting constraints are summarized in Table \ref{tab:cosmo_params} and illustrated in Figure \ref{fig1}.

We find that the results from the Planck-$EB$ (with Planck+DESI+PantheonPlus) dataset differ markedly from our previous analysis \cite{Kochappan:2024jyf} (which utilized Planck-$EB$, Planck, BOSS \cite{BOSS:2016wmc}, and SH0ES \cite{Riess:2021jrx}). Specifically, the current analysis yields a significantly higher value for $H_0$, accompanied by an increase in both the EDE fraction $f_{\rm EDE}$. { These shifts imply that, compared to our previous analysis, the inclusion of 
DESI and Pantheon+ favors a larger cosmic birefringence coupling $gM_\mathrm{Pl}$ and a higher EDE fraction $f_\mathrm{EDE}$.} As shown in Table \ref{tab:cosmo_params}, both the Planck-$EB$ and ACT-$EB$ analyses yield consistent and compelling constraints. A salient feature of our results is that both datasets prefer a relatively high Chern--Simons coupling constant $g M_{\mathrm{Pl}}$ accompanied by a substantial fraction of EDE, $f_{\rm EDE}$. Specifically, we find $f_{\rm EDE} = 0.215^{+0.083}_{-0.049}$ for the Planck-$EB$ case and $f_{\rm EDE} = 0.232^{+0.074}_{-0.047}$ for the ACT-$EB$ case. Theoretically, this enhanced EDE component acts to reduce the comoving sound horizon at recombination; consequently, a higher Hubble constant is required to maintain consistency with the observed angular scale of the acoustic peaks. This mechanism leads to derived $H_0$ values of $76.2^{+3.1}_{-2.5}$ km s$^{-1}$ Mpc$^{-1}$ and $76.9^{+2.9}_{-2.5}$ km s$^{-1}$ Mpc$^{-1}$ for the Planck-$EB$ and ACT-$EB$ combinations, respectively. These results are in excellent agreement with the local distance ladder measurement from the SH0ES team ($H_0 = 73.04 \pm 1.04$ km s$^{-1}$ Mpc$^{-1}$), demonstrating that the cosmic birefringence and EDE provide a robust resolution to the Hubble tension. { To quantitatively assess the agreement between our results and those of SH0ES \cite{Riess:2021jrx},
we compute the Gaussian tension (GT) between our best-fit $H_0$ posterior and the
SH0ES measurement $H_0^\mathrm{SH0ES} = 73.04 \pm 1.04$. The Gaussian tension is defined as
\begin{equation}
    \sigma_\mathrm{GT}
    = \frac{\left|\mu_1 - \mu_2\right|}{\sqrt{\sigma_1^2 + \sigma_2^2}},
\end{equation}
where $\mu_1$ and $\sigma_1$ denote the mean and symmetrised $1\sigma$ uncertainty of
our $H_0$ posterior with $\sigma_1 $ equal to $ (\sigma_+ + \sigma_-)/2$ in Table \ref{tab:cosmo_params}. $\mu_2$ and $\sigma_2$ are the corresponding SH0ES values.
For the Planck-$EB$ combination,
we obtain
$\sigma_\mathrm{GT}^\mathrm{Planck\text{-}EB}
    = \frac{|76.2 - 73.04|}{\sqrt{2.8^2 + 1.04^2}}
    \approx 1.06\sigma.$
For the ACT-$EB$ combination,
$\sigma_\mathrm{GT}^\mathrm{ACT\text{-}EB}
    = \frac{|76.9 - 73.04|}{\sqrt{2.7^2 + 1.04^2}}
    \approx 1.33\sigma.$
Both values are well below the conventional $2\sigma$ threshold; this offset is not statistically significant given the width of our posterior. 

{For the Planck $low-\ell$ EE likelihood in Table \ref{tab:chi2} , the expected $\chi^2$ for our Planck-EB (with Planck+DESI+PantheonPlus) is consistent with the result in Table V of Ref.~ \cite{Hill:2020osr}. 
The ACT-EB (with Planck+DESI+PantheonPlus) combination, however, gives $\chi^2$ = 462.93, an increase of 65 over the expected baseline.}
}

The consistency of these findings is further illustrated in Figure \ref{fig1}, which presents the marginalized one- and two-dimensional posterior distributions for the key parameters. The substantial overlap between the Planck-$EB$ (blue color) and ACT-$EB$ (purple color) contours indicates that the preference for this birefringence-EDE model is robust against the choice of CMB polarization scales. In the 2D contour panels, we observe a clear positive degeneracy between the critical redshift $\log_{10} z_c$ and the coupling constant $g M_{\mathrm{Pl}}$. This correlation suggests that as the EDE injection shifts to earlier times (higher $z_c$), the scalar field's evolution during the recombination window becomes less efficient at sourcing birefringence, thereby requiring a larger coupling constant $g M_{\mathrm{Pl}}$ to reproduce the observed rotation signal. Furthermore, a strong positive degeneracy is observed between $f_{\rm EDE}$ and $H_0$, confirming that the preference for a higher Hubble constant is directly driven by the increased EDE density, which modifies the pre-recombination expansion history.

{  
The parameter $\tau_\mathrm{reio}$ plays a pivotal role in connecting the earliest astrophysical processes in the Universe to CMB observations. Its magnitude is determined by the total production and evolutionary history of ionizing photons during the epoch of reionization. A larger $\tau_\mathrm{reio}$ implies that neutral hydrogen in the interstellar and intergalactic media was ionized more efficiently, that reionization commenced at earlier times, and that the process proceeded more rapidly overall. Such a scenario requires a sufficiently abundant population of high-energy ultraviolet photons, which in turn suggests more rapid and prolific formation of early galaxies, particularly low-mass systems, capable of supplying the necessary ionizing radiation. In the present analysis, both data combinations yield $\tau_\mathrm{reio} \approx 0.135$--$0.136$, noticeably higher than the Planck 2018 baseline value of $\tau_\mathrm{reio} = 0.054 \pm 0.007$. We note that this baseline value is obtained from a $TT$+$TE$+$EE$-only analysis, and that incorporating additional low-redshift datasets is known to shift $\tau_\mathrm{reio}$ systematically upward, to 
$0.0581$--$0.0626$ \cite{Hill:2020osr, Poulin:2018cxd}. This trend continues in our own sequence of analyses: fitting Planck-$TT$+$EE$+$EB$+SH0ES+BOSS in our previous work \cite{Kochappan:2024jyf} already yielded $\tau_\mathrm{reio} = 0.0711$, and replacing BOSS and SH0ES with PantheonPlus and DESI DR2 in the present analysis raises this value further to $0.135$. We attribute this increase primarily to the tighter constraints on the late-time expansion history imposed by PantheonPlus. 
{The ionizing-photon output of observed high-$z$ galaxies does not
require reionization to extend this far. Quasar/Lyman-$\alpha$ forest
data show the IGM is already highly ionized by $z \approx 5.3$--$6$,
and kSZ measurements favor $\tau_{\rm reio} \approx 0.05$--$0.06$ \cite{Pagano:2019tci, Bosman:2021oom}. We
therefore flag $\tau_{\rm reio} =0.135$ as a limitation of the
model, alongside the $S_8$ tension, and note that it merits further
investigation in future work.}}

\begin{figure}
\centering
\includegraphics[width=0.75\linewidth]{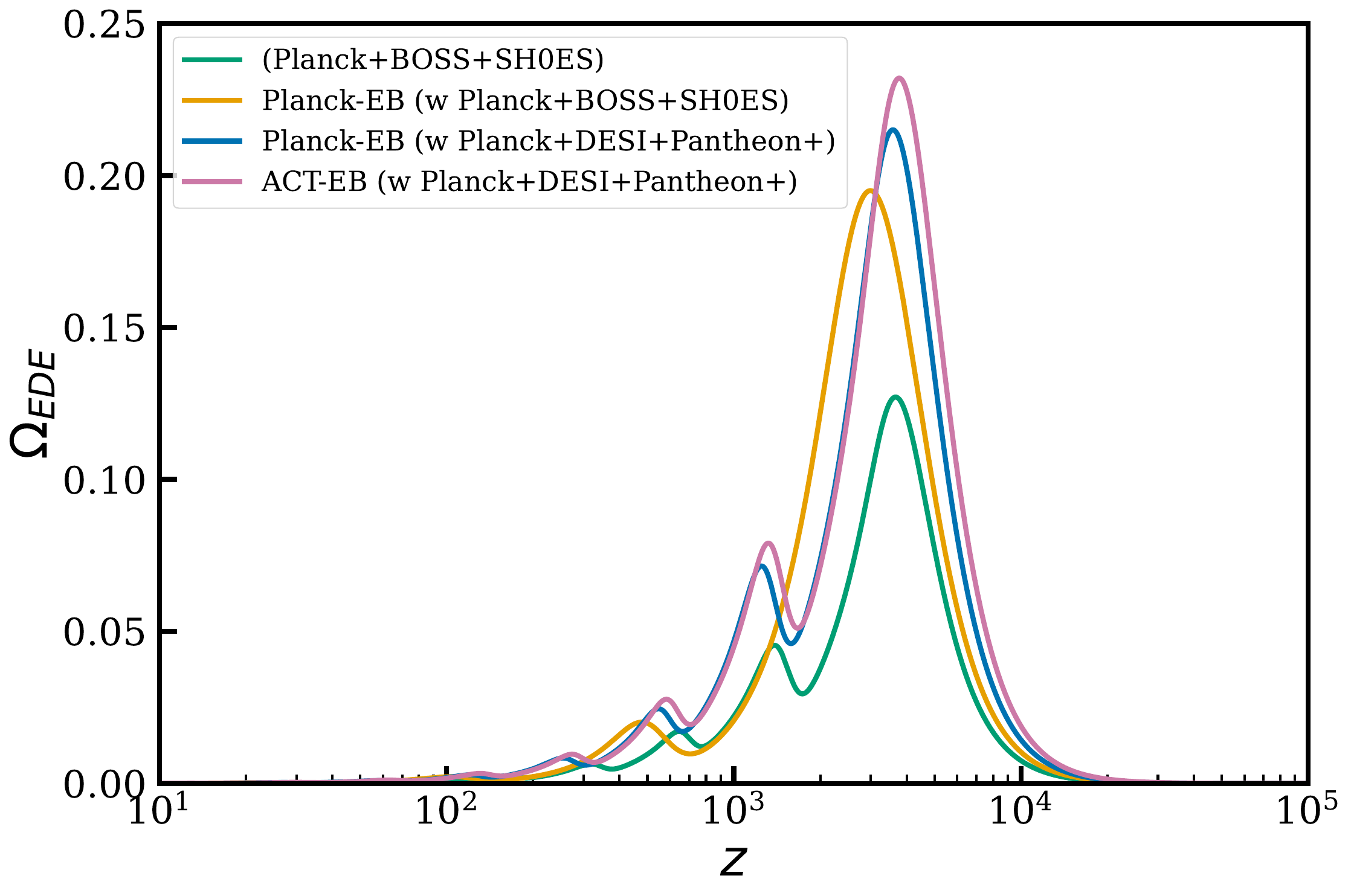}
\caption{\label{fig:rho} The ratio of the EDE density $\Omega_{\mathrm{EDE}}$ as a function of redshift $z$.
{ 
For the four data combinations introduced in the text: 
green line from Ref.~\cite{Eskilt:2023nxm}; yellow line from  Ref.~\cite{Kochappan:2024jyf}; blue and purple line are the best-fit result from this work. The corresponding initial 
conditions $\theta_i$ are $2.768$, $1.89$, 
    $2.7$, $2.7$ in green, yellow, blue and purple lines, respectively.
The $f_\mathrm{EDE}$ are $0.1271$, $0.1950$, $0.215$, $0.232$ in green, yellow, blue and purple lines, respectively. And $H_0$ values are $69.71$, 
    $72.03$, $76.2$, 
    $76.9$ in green, yellow, blue, and purple lines, respectively.}
}
\end{figure}

We next compare the results obtained by Ref.~\cite{Eskilt:2023nxm}, our previous \cite{Kochappan:2024jyf} publication, and the new constraints derived from the data combinations analyzed in this work. Figure \ref{fig:rho} illustrates the fraction of EDE relative to the total cosmic energy density under different data fits.
Here we use $\Omega_{EDE}=\rho_{EDE}/\rho_{tot}$.
The green line corresponds to the results presented by Ref. \cite{Eskilt:2023nxm}, obtained by fitting Planck $TT$, $TE$, and $EE$ data together with BOSS and SH0ES, under the assumption that cosmic birefringence is absent ($gM_\mathrm{Pl}=0$). { Comparing with Ref.~\cite{Kochappan:2024jyf}, the BOSS and SH0ES datasets have  been replaced by DESI DR2 and PantheonPlus, and $\theta_i$ is now fixed to  $2.7$ rather than varied freely.} Among the four cases shown, this scenario yields the smallest EDE contribution.
The yellow line represents the constraints from our earlier analysis \cite{Kochappan:2024jyf}, in which Planck $EB$ data were incorporated to constrain the cosmic birefringence coupling parameter $gM_\mathrm{Pl}$. Compared with the birefringence-free case, the inclusion of $EB$ polarization data clearly favors a larger fraction of EDE.

The blue and purple lines correspond to the new results obtained by jointly fitting an EDE model with cosmic birefringence using the Planck+DESI+PantheonPlus baseline, further augmented by Planck-$EB$ and ACT-$EB$ data, respectively. In these analyses, a total of 9 free parameters are varied simultaneously as shown in Table \ref{tab:cosmo_params}. The updated data combinations allow higher values of $\Omega_{\rm EDE}$, with the ACT-$EB$-enhanced fit (purple line) yielding a systematically larger EDE fraction than that inferred from Planck-$EB$ alone.
In all cases, the EDE component reaches its maximum contribution around the epoch when the matter and radiation energy densities are comparable, which defines the parameter $f_{\rm EDE}$. After the CMB epoch, the EDE fraction rapidly decreases and becomes negligible by the present time.

\begin{figure}
\centering
\includegraphics[width=0.75\linewidth]{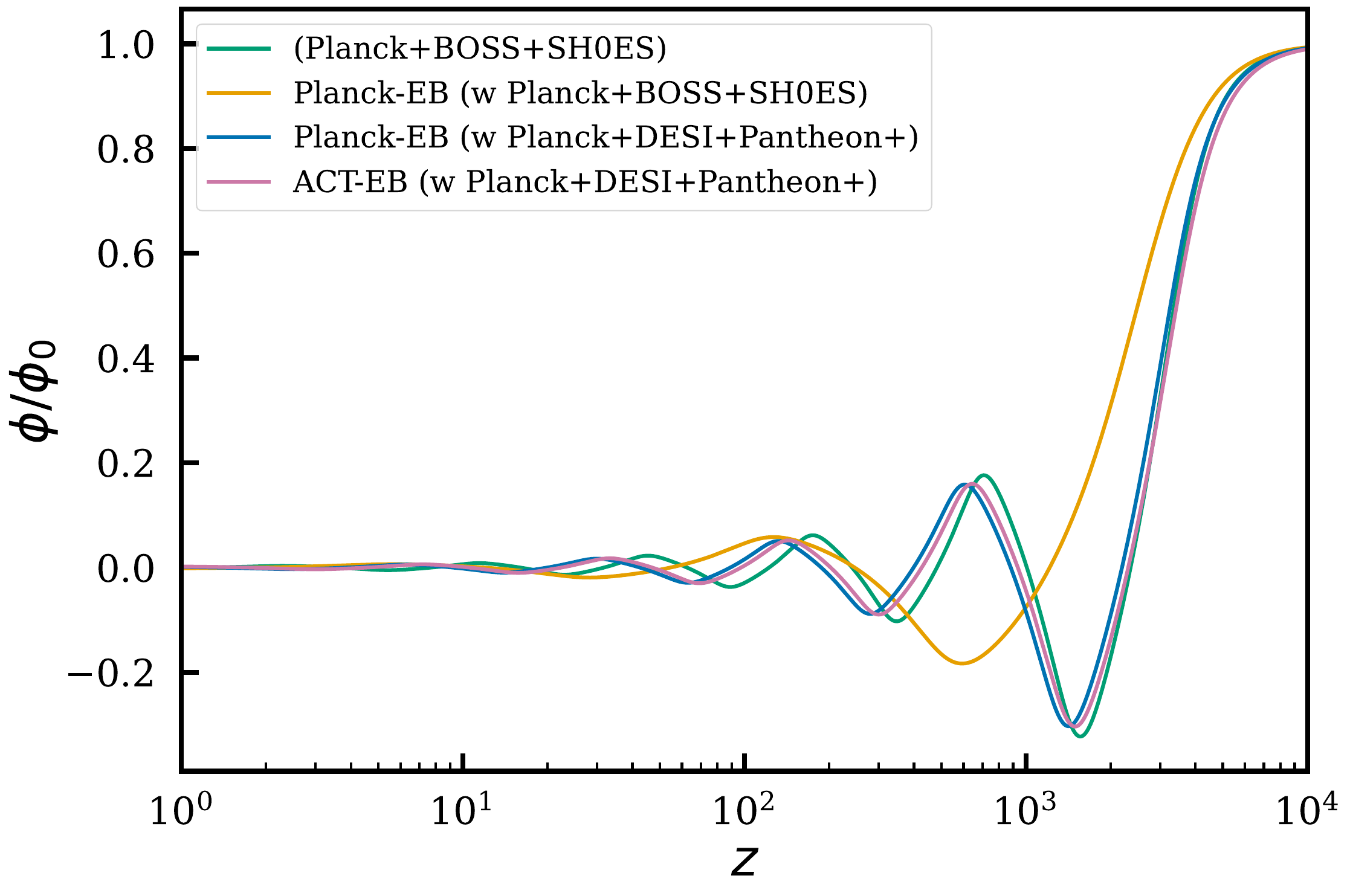}
\caption{\label{fig:theta}  The ratio of the EDE scalar field 
$\phi$ to its initial value $\phi_0$
 as a function of redshift $z$. The corresponding initial 
conditions $\theta_i$ are $2.768$, $1.89$, 
    $2.7$, $2.7$ in green, yellow, blue and purple lines, respectively. }
\end{figure}

Figure~\ref{fig:theta} illustrates the oscillatory behavior of the scalar-field potential under four group datasets. Since the field amplitude satisfies $\phi = \theta f$ and the best-fit values of $\theta$ differ among the various data combinations, we normalize the field as $\phi / \phi_0$ in order to compare the redshift evolution of the potential oscillations on equal footing.
During the epoch of recombination, the potential transitions from a slow-roll regime into a phase of damped oscillations. As shown in the plot, the oscillation frequency depends sensitively on the value of $\theta_i$. In particular, the yellow line, which corresponds to the smallest best-fit $\theta_i$-even below 2-exhibits the slowest oscillatory behavior among all cases.
After the matter-radiation equivalent epoch, the oscillations of the potential are rapidly damped. By the present time, the EDE field has effectively settled, and no residual oscillatory contribution from EDE remains in the late Universe.

\begin{figure}
\centering
\includegraphics[width=0.75\linewidth]{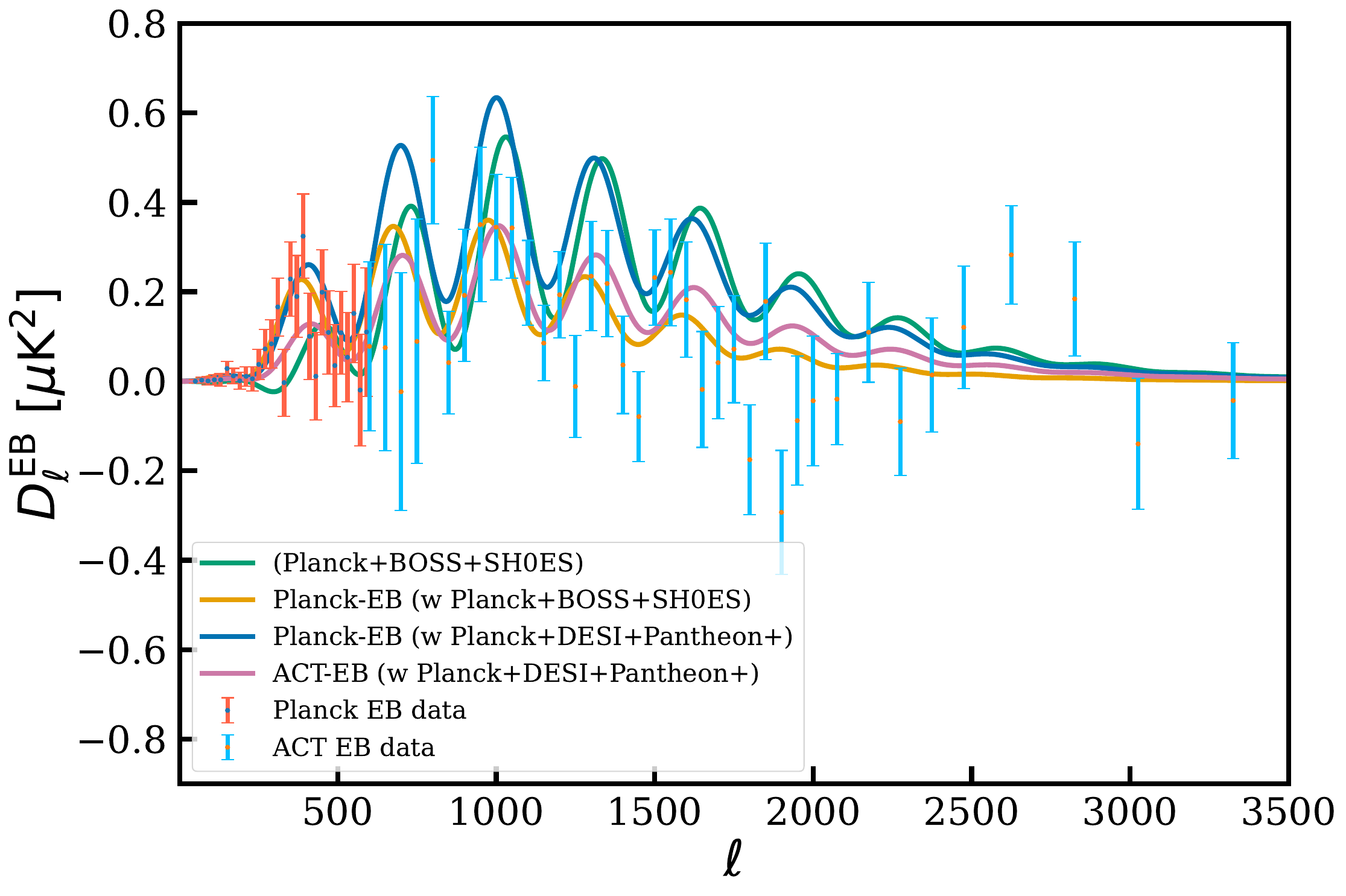}
\caption{\label{fig:Dleb} The CMB $D_\ell^{EB}$ power spectra with the best-fit result from four groups' data. The Planck-$EB$ data and ACT-$EB$ data points with errorbars are shown in tomato and light-blue colors, while the plot excludes the overlapping multipole range of the Planck-$EB$ data. }
\end{figure}

\begin{figure}
\centering
\begin{tabular}{cc}
\includegraphics[width=0.45\linewidth]{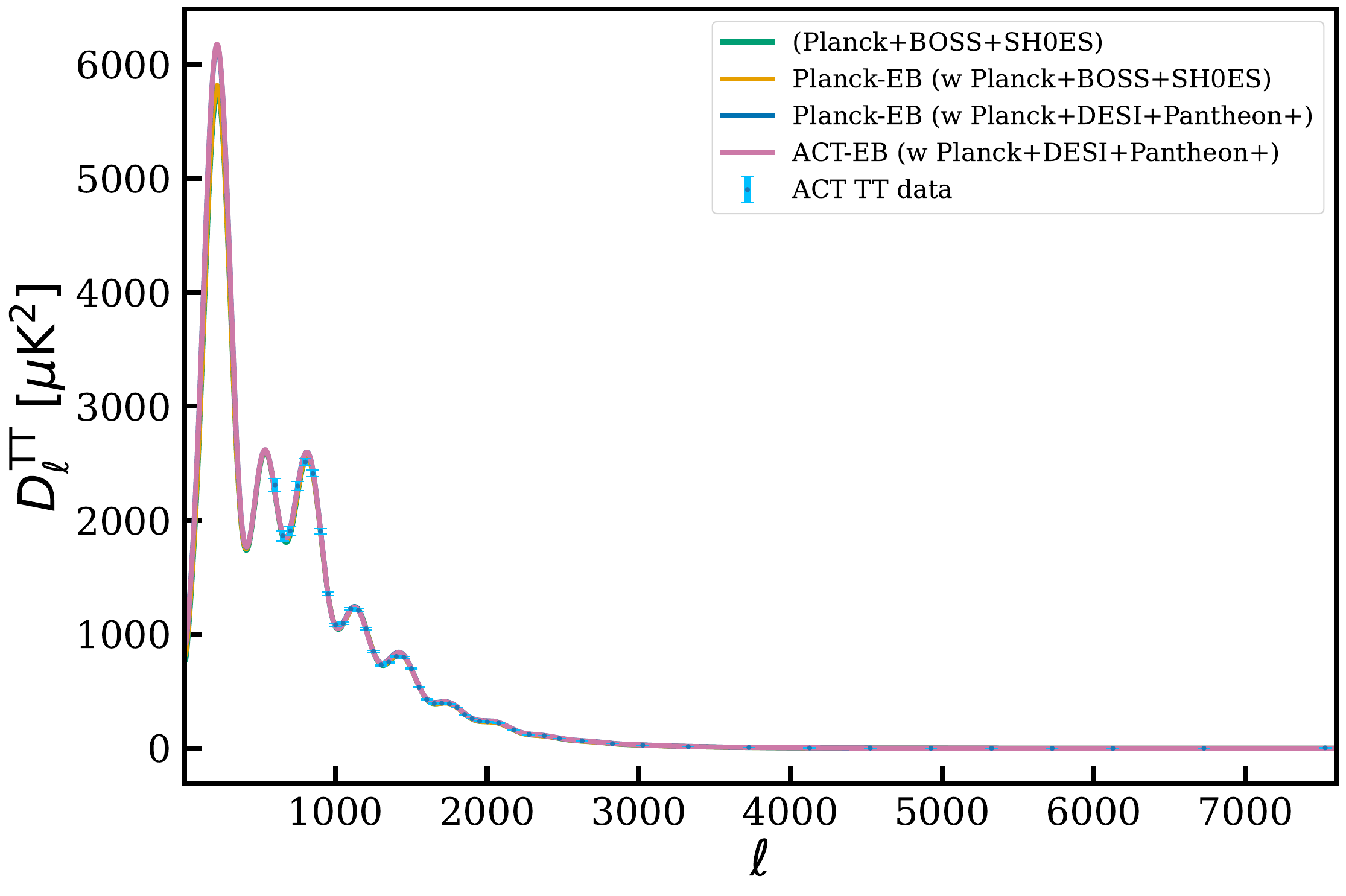}&
\includegraphics[width=0.45\linewidth]{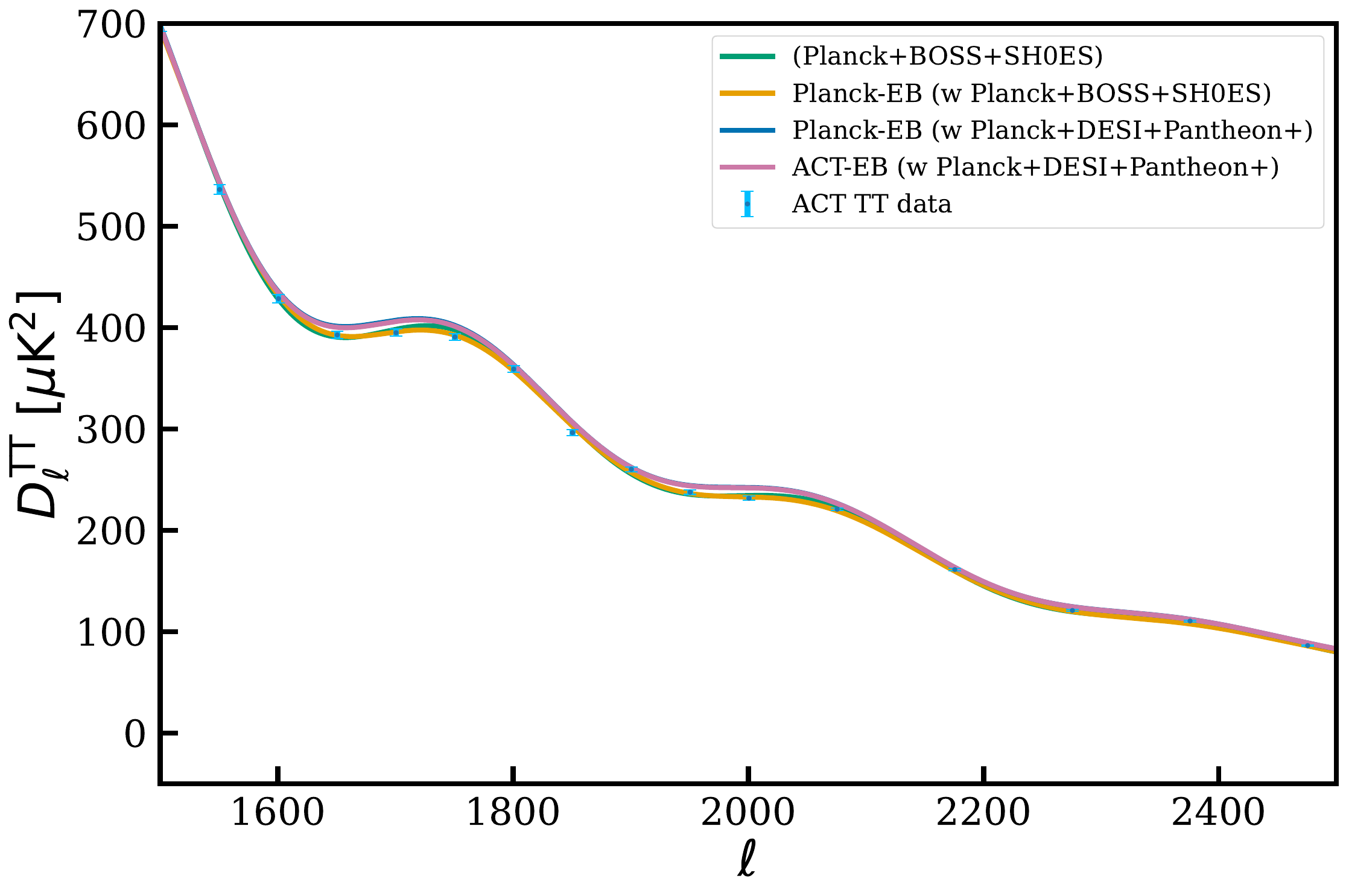}\\
(a)  & (b)   \\
\end{tabular}
\caption{\label{fig:Dltt} {  The CMB $D_\ell^{TT}$ power spectra with the best-fit result from four groups' data. The light-blue dataset from ACT DR6 temperature power spectrum. The panel (a) shows the range of $\ell$ from 2 to 7600, while the  panel  (b) provides a zoomed-in view of  $\ell$ from 1500 to 2500.} }
\end{figure}

\begin{table}[htbp]
\centering
\caption{Breakdown of the best-fit $\chi^2$ values for different datasets.}
\label{tab:chi2}
\begin{tabular}{lcc}
\hline\hline
Dataset & Planck-$EB$ & ACT-$EB$ \\
\hline
Planck low-$\ell$ TT & 30.11 & 37.31 \\
Planck low-$\ell$ EE & 398.64 & 462.93 \\
High-$\ell$ EE & 388.37 & 383.13 \\
High-$\ell$ TT & 432.30 & 424.09 \\
Planck lensing & 302.56 & 133.05 \\
EB likelihood & 197.76 & 119.35 \\
DESI DR2 BAO & 42.44 & 16.90 \\
PantheonPlus & 1403.12 & 1405.85 \\
\hline
Total $\chi^2$ & 3195.31 & 2982.60 \\
\hline\hline
\end{tabular}
\end{table}

 We compare the differences between the theoretical predictions and the observational measurements of the $EB$ power spectra for four different data combinations in Figure \ref{fig:Dleb}. The tomato errorbar and data points correspond to the Planck $EB$ measurements, while the light-blue points represent the ACT-$EB$ data. The multipole coverage of the two experiments differs substantially: Planck probes the range $50 \leq \ell \leq 1500$, whereas ACT extends from $\ell = 600$ up to $\ell = 3500$. In the figure, we display the full ACT multipole range together with the portion of the Planck data that does not overlap with ACT.
The green line shows the theoretical prediction obtained by first fitting the EDE component and subsequently introducing cosmic birefringence as a secondary effect, with the Chern-Simons coupling parameter $gM_\mathrm{Pl}$ constrained using Planck-$EB$ data alone. At present, the observational precision at very low and very high multipoles remains limited, while the intermediate $\ell$ range is generally regarded as the most accurately measured region of the CMB.

Notably, both the blue and green lines overshoot the ACT-$EB$ measurements more than one-sigma errorbar at the two peaks located in the multipole range around $\ell \simeq 1000$-1500. 
{  To quantify this comparison, we computed the $\chi^2$ and PTE of each  best-fit model against the 38 ACT-$EB$ data points. The Planck-$EB$ (with Planck+DESI+Pantheon+) curve yields $\chi^2 = 117.6$ (PTE $=2.49\times 10^{-10}$),  confirming the visible overshoot, while the ACT-$EB$ (with Planck+DESI+Pantheon+) curve yields $\chi^2 = 53.4$ (PTE $= 0.039$), a substantially improved but still marginal fit.}
The $EB$ power spectra result of the green line comes from only fitting $gM_\mathrm{Pl}$ in Planck-$EB$ and other EDE parameters fixed in the Planck+BOSS+SH0ES data combination.
 In contrast, the yellow line corresponds to a simultaneous fit of all ten free parameters, significantly alleviating this tension and providing a much improved agreement with the observations.
The dark-blue line represents the results obtained when fitting all parameters simultaneously, the BAO dataset replaced by the latest DESI measurements, and the SH0ES supernova sample upgraded to the larger PantheonPlus compilation. 
In this case, since the Planck-$EB$ data have larger uncertainties in the multipole range $\simeq 1000$-1500 than ACT-$EB$, the predicted birefringence power spectrum amplitude becomes too large and exceeds the ACT-$EB$ error bars. 
By contrast, for the same underlying data combination, the purple line demonstrates that replacing Planck-$EB$ with ACT-$EB$ effectively suppresses the height of the $D_\ell^{EB}$ peaks, yielding a better match to the most precise measurements in the intermediate multipoles regime.
{ To further assess the consistency of the large EDE fraction $f_\mathrm{EDE}$ with small-scale CMB observations, we compare our best-fit theoretical predictions with the ACT DR6 temperature power spectrum. As shown in Figure \ref{fig:Dltt} (a), all four data combinations remain broadly consistent with the ACT $TT$ measurements across the full multipole range. 
Panel (b), which shows the zoomed-in region $1500 \lesssim \ell \lesssim 2500$,  reveals that the two combinations incorporating DESI and PantheonPlus predict $TT$ power slightly above the ACT DR6 data points near $ 1700$ and $2000$. This confirms that the large $f_\mathrm{EDE}$ favored by our joint analysis is not in significant conflict with the small-scale $TT$ spectrum.
Meanwhile, the yellow line lies closer to the ACT $TT$ observations than the green line, consistent with our earlier finding that incorporating 
the $EB$ data into the global fit yields tighter and more self-consistent constraints across all multipole ranges \cite{Kochappan:2024jyf}.}

\begin{figure}
\centering
\begin{tabular}{cc}
\includegraphics[width=0.49\linewidth]{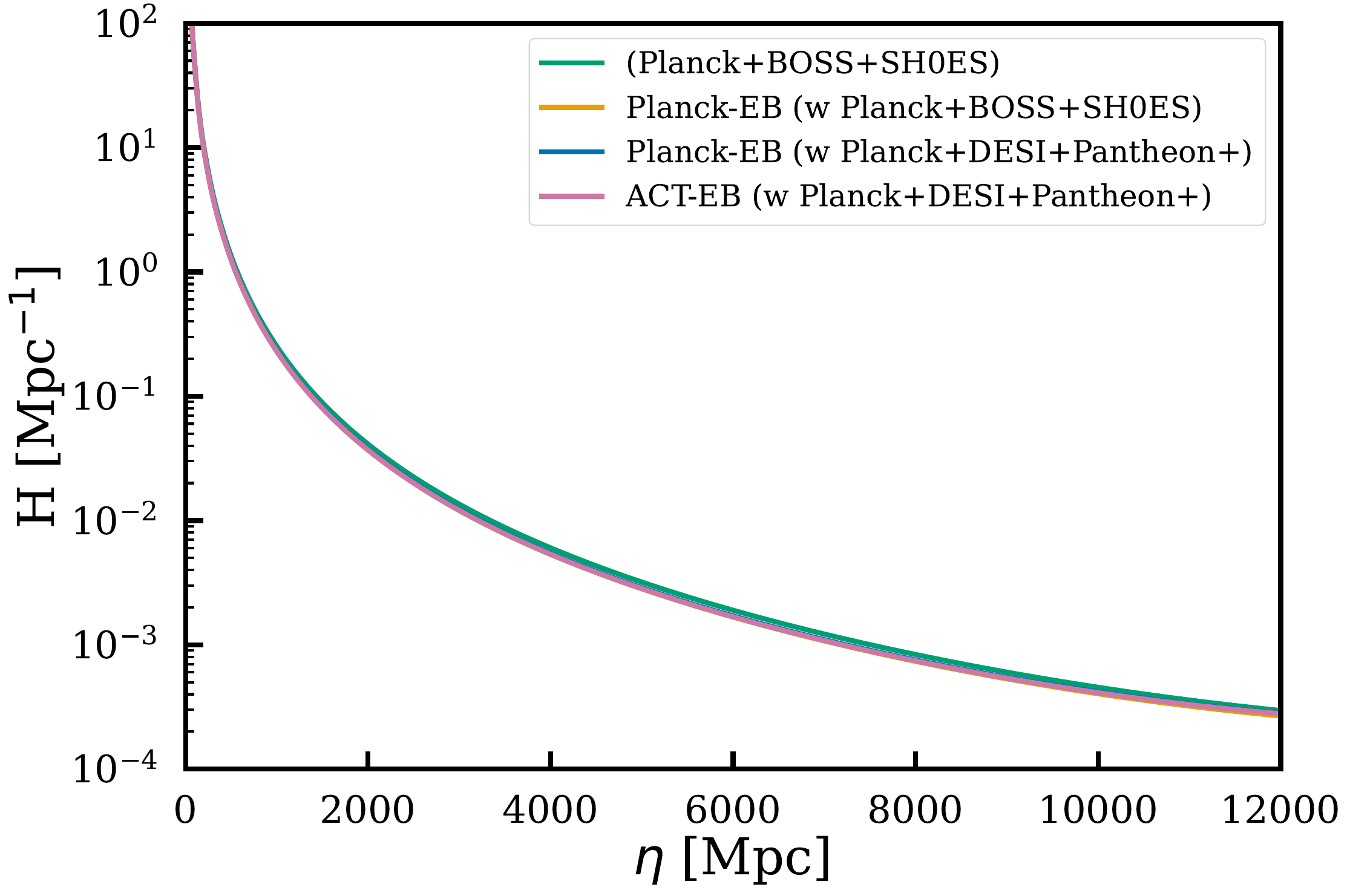}&
\includegraphics[width=0.49\linewidth]{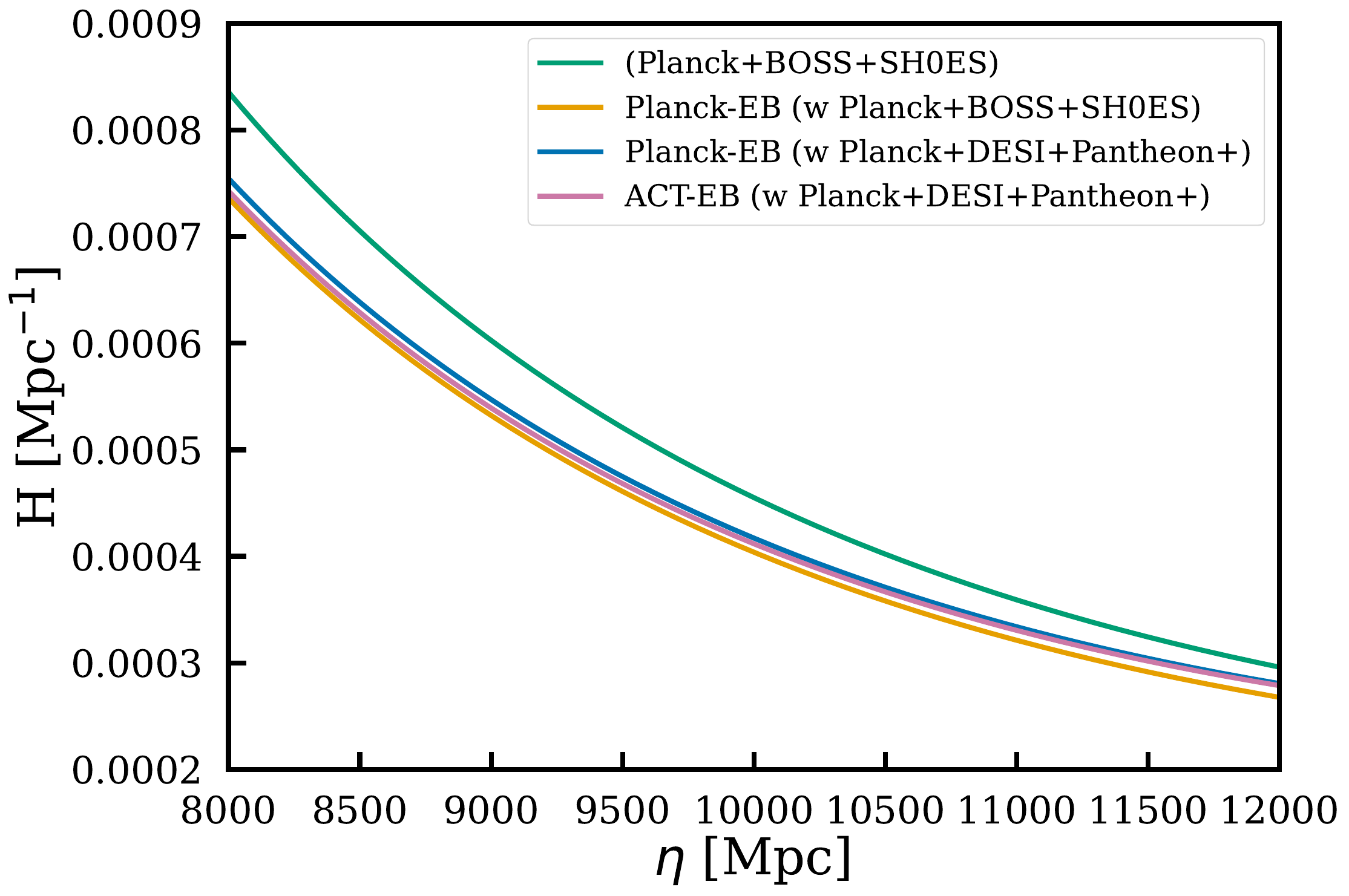}\\
(a)  & (b)   \\
\end{tabular}
\caption{\label{fig:Htau} The evolution of the Hubble parameter as a function of conformal time $\eta$ for the four data combinations. The panel  (a) shows the range of $\eta$ from 0 to 12000, while the panel (b) provides a zoomed-in view of  $\eta$ from 8000 to 12000. }
\end{figure}

Figure \ref{fig:Htau} illustrates the evolution of the Hubble parameter as a function of conformal time $\eta$ across the four distinct data combinations. The left panel displays the overall decreasing trend of the Hubble parameter over the range $\eta \in [0, 12000]$, while the right panel presents a zoomed-in view of the interval $\eta \in [8000, 12000]$. This magnified region allows us to resolve the relative ordering and subtle differences among the four best-fit solutions. { 
Meanwhile, the elevated $\Omega_c h^2$ values reported in Table~\ref{tab:cosmo_params}, driven by the large EDE fraction, have direct implications for the growth of cosmic structures at late times. The increased matter density, together with the enhanced amplitude of matter fluctuations $\sigma_8$, results in relatively high values of $S_8 \equiv \sigma_8\sqrt{\Omega_m/0.3}$, namely $S_8 = 0.964$ for the Planck-$EB$ (Planck+DESI+PantheonPlus) dataset and $S_8 = 0.967$ for the ACT-$EB$ (Planck+DESI+PantheonPlus) dataset. Both data combinations $S_8$ are in tension with the KiDS-Legacy measurement $S_8 = 0.815^{+0.016}_{-0.021}$ \cite{Wright:2025xka} at the $5.1\sigma$ (Planck-EB) and $5.2\sigma$ (ACT-EB) level, and with the DESY6 result $S_8 = 0.789 \pm 0.012$ \cite{DES:2026fyc} at $6.7\sigma$ and $6.9\sigma$, respectively. 
This discrepancy is a well-known challenge for EDE models with large $f_{\rm EDE}$ \cite{Hill:2020osr,Smith:2019ihp}, as the larger early dark energy contribution generally requires a higher matter density and consequently predicts enhanced structure growth at late times. Since no weak-lensing likelihood is included in the present analysis, the resulting $S_8$ tension is not directly constrained. A comprehensive joint analysis incorporating weak-lensing data is therefore deferred to future work.}

\begin{figure}
\centering
\begin{tabular}{ccc}
\includegraphics[width=0.33\linewidth]{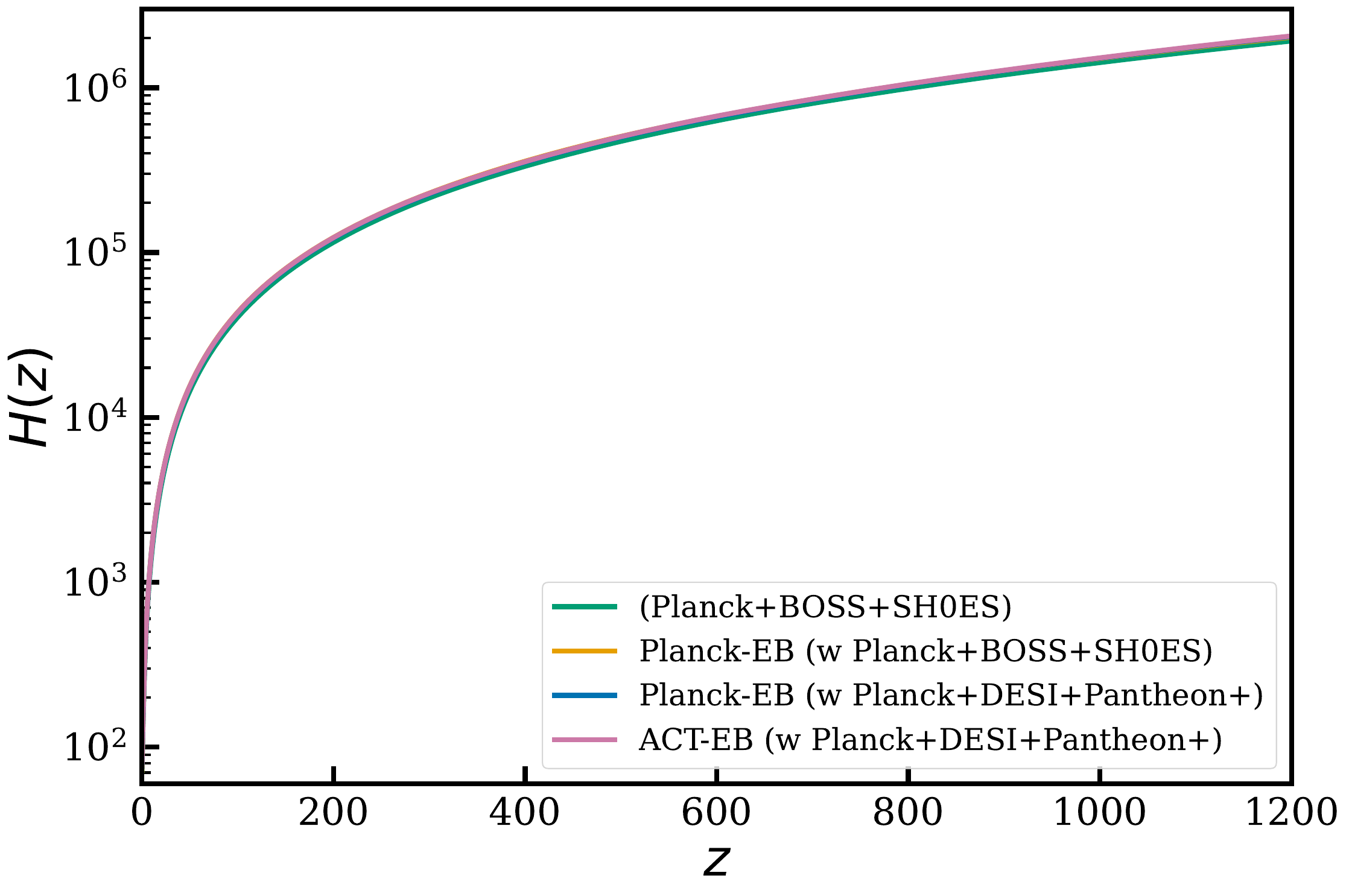}&
\includegraphics[width=0.33\linewidth]{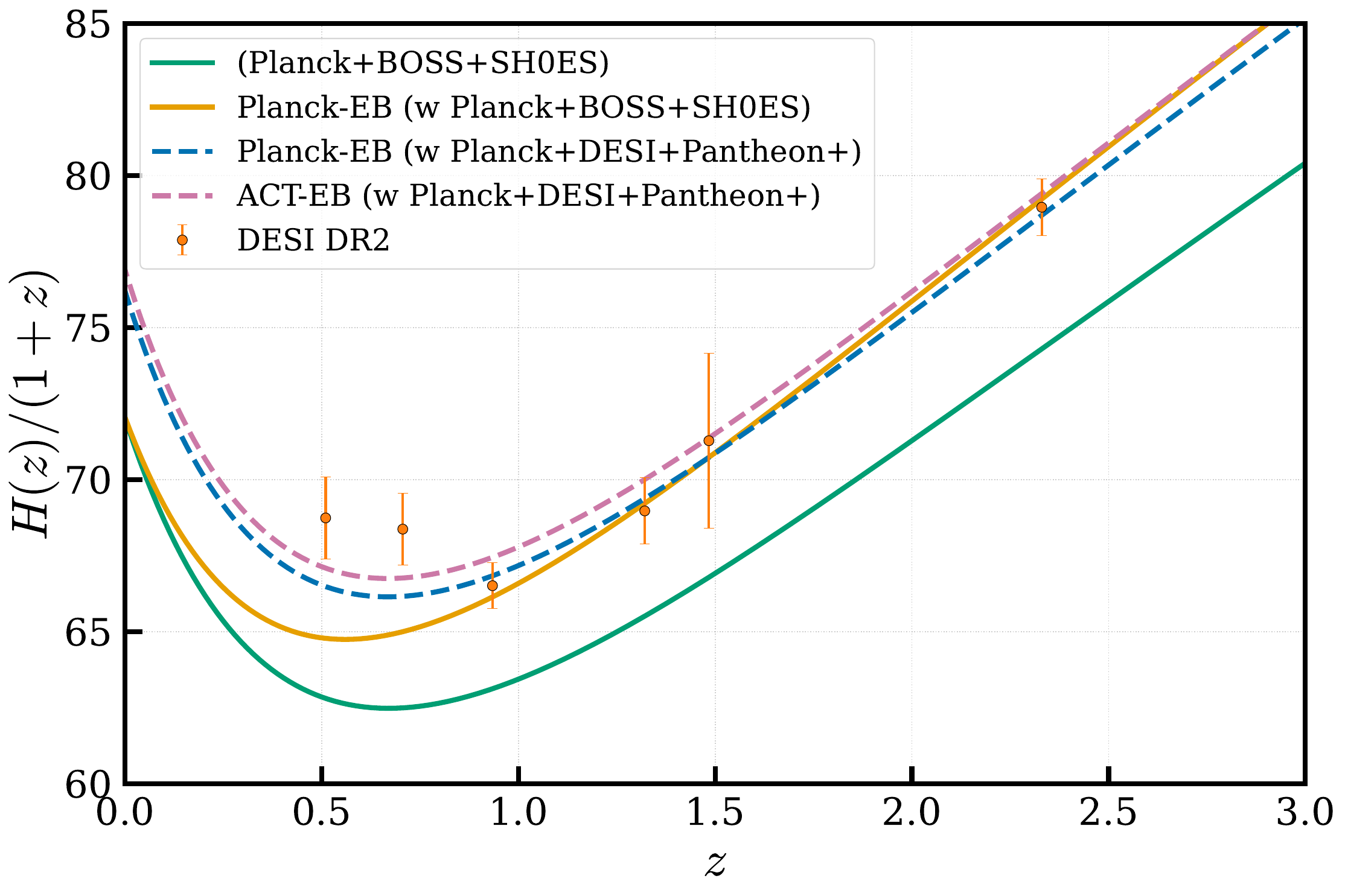}&
\includegraphics[width=0.33\linewidth]{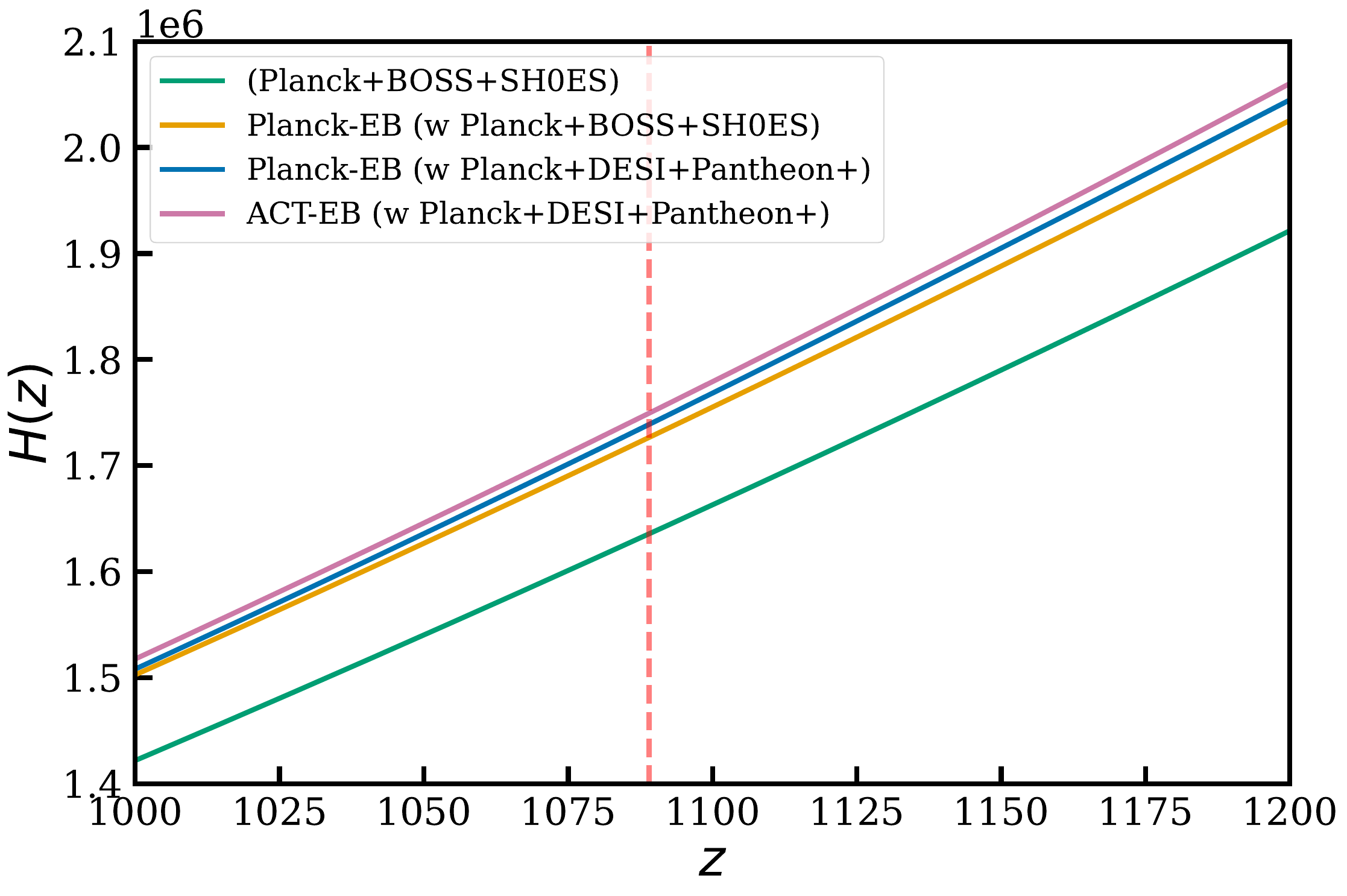}\\
(a)  & (b) & (c)  \\
\end{tabular}
\caption{\label{fig:Hz}  The evolution of the Hubble parameter with redshift inferred from these four data combinations. Figures (a), (b), and (c) show the redshift range from 0 to 1200, the local Universe $z\in [0, 3]$, and the early Universe $z\in [1000, 1200]$, respectively.}
\end{figure}

In Figure \ref{fig:Hz}, we focus more closely on comparing the redshift evolution of the Hubble parameter inferred from these four data combinations. Figure~\ref{fig:Hz} (a) shows $H(z)$ over the redshift range $0 \leq z \leq 1200$, where all four datasets predict broadly consistent expansion histories. In figure~\ref{fig:Hz} (b), we zoom in on the late-time Universe, $0 \leq z \leq 3$, to highlight differences in the expansion rate. The green line corresponding to the EDE fit did not include the cosmic birefringence effect, which shows the lowest late-time expansion rate. Once cosmic birefringence is taken into account, the other three fits predict systematically higher expansion rates at late times.
The blue and purple dashed lines incorporate the latest DESI and PantheonPlus datasets, leading to an enhancement of $H(z)$ at low redshifts, particularly for $z < 1$, while their impact at higher redshifts remains comparatively small and close to the yellow line by Planck-$EB$ with Planck+BOSS+SH0ES data result.
Figure~\ref{fig:Hz} (c) shows the behavior of the Hubble parameter in the early Universe, especially the CMB epoch. The red dashed vertical line marks the time of CMB. The birefringence-free case (green line) again exhibits the smallest value of $H(z)$ at recombination. Meanwhile, all three data combinations that include cosmic birefringence in the MCMC analysis predict larger values of $H(z)$. Among these, the Hubble parameter at the CMB epoch decreases slightly in sequence for the ACT-$EB$ (with Planck+DESI+PantheonPlus), Planck-$EB$ (with Planck+DESI+PantheonPlus), and Planck-$EB$+Planck+BOSS+SH0ES fits.

\section{Summary}
\label{sec:4}

The phenomenon of cosmic birefringence, particularly its effects on CMB photons, is becoming a prominent topic as increasingly precise CMB detector experiments continue to validate it. In this work, we employ a combination of Planck, DESI, PantheonPlus, and ACT-$EB$ polarization data,  used with MCMC techniques, to investigate the interplay among cosmological parameters in an EDE scenario that interacts with photons in Chern-Simons coupling to produce cosmic birefringence.

By comparing the MCMC constraint results from Planck-$EB$ and ACT-$EB$, each combined with data from Planck+DESI+PantheonPlus for the EDE model incorporating cosmic birefringence, we obtained the constraint results for nine parameters and their corresponding $\chi^2$ value.
We find that incorporating birefringence effects in the early Universe, especially the EDE model, can significantly enhance the inferred value of the Hubble constant $H_0$ and EDE's highest energy density $f_\mathrm{EDE}$. This naturally leads to an increased EDE fraction during recombination. When late-time datasets such as DESI and PantheonPlus are included, the other parameters were also constrained, yielding more precise results.

We also compared the redshift evolution of $\Omega_{EDE}$  in the EDE model across different datasets, including Planck+BOSS+SH0ES, Planck-$EB$ with Planck+BOSS+SH0ES, Planck-$EB$ with Planck+DESI+PantheonPlus, and ACT-$EB$ with Planck + DESI + PantheonPlus, based on previous studies. Additionally, comparisons were made regarding the oscillation of the potential, the power spectra
$D_\ell^{EB}$ under the four best-fit scenarios against results from Planck-$EB$ and ACT-$EB$. Owing to the smaller error bars in the $D_\ell^{EB}$ measurements provided by ACT-$EB$, the resulting constraints on the birefringence coupling parameter $gM_\mathrm{Pl}$ are tighter and yield systematically smaller values than those inferred from Planck-$EB$. This suggests that analyses relying solely on Planck-$EB$ data may no longer be fully compatible with the current ACT observations. The evolution of the Hubble parameter with conformal time and redshift with different datasets is also compared in Figures \ref{fig:Htau} and \ref{fig:Hz}.
Forthcoming CMB polarization experiments such as LiteBIRD and AliCPT are expected to deliver substantially improved sensitivity. These observations will enable more stringent and robust constraints on cosmic birefringence and its coupling to EDE, offering a promising avenue for testing such extensions of the standard cosmological model. {The cosmic birefringence effect induced by late-time dark energy is also a worthwhile topic to explore in future studies, and these new experiments potentially help us distinguish phenomena associated with late-time dynamical dark energy.}

\section*{Acknowledgements}

This work was supported by the National Natural Science Foundation of China (Grants Nos. 12533001 and 12473001), the National SKA Program of China (Grants Nos. 2022SKA0110200 and 2022SKA0110203), the China Manned Space Program (Grant No. CMS-CSST-2025-A02), and the National 111 Project (Grant No. B16009). 
L.Yin was supported by the Natural Science Foundation of Shanghai 24ZR1424600.

\section*{Appendix}

\begin{figure}[H]
    \centering
    \includegraphics[width=0.7\linewidth]{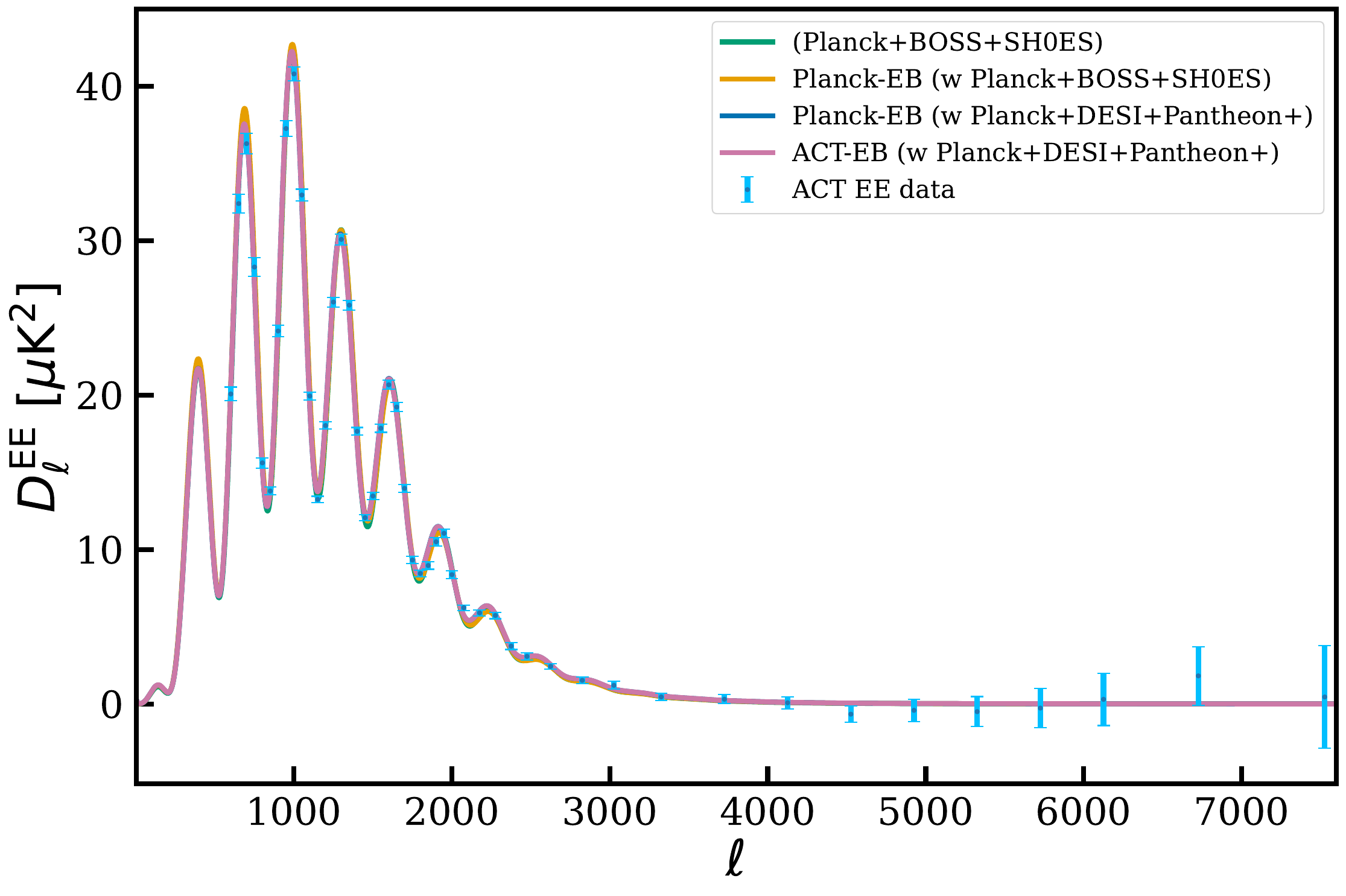}
    \caption{The CMB $D_\ell^{EE}$ power spectra with the best-fit result from four groups’ data. }
    \label{fig:ee}
\end{figure}

We add the CMB-$EE$ power spectra with the best-fit result from four groups’ data comparing with ACT-$EE$ data points, as shown in Figure~\ref{fig:ee}.

\end{document}